# Comparing Bayesian Dynamic Borrowing and Synthetic Control Methods: Apples to Oranges?

Nicole Cizauskas, Foteini Strimenopoulou, Svetlana S. Cherlin, James M. S. Wason.

## Abstract

*Introduction*

Bayesian dynamic borrowing (BDB) and synthetic control methods (SCM) are strategies to incorporate historical data in clinical trial design when control recruitment is a challenge. While both methods aim to improve efficiency, they differ in conceptual approach: BDB adjusts borrowing based on prior-data conflict, whereas SCM synthesizes individual-level control outcomes without explicitly modelling variability between studies Comparing these methods has been limited due to their differences in application, measurement metrics, and the general difficulty of comparing across Bayesian and frequentist methods. This study aims to conduct a comparison BDB (using meta-analytic predictive prior (MAP)) and SCM using a case study of Pediatric Atopic Dermatitis.

*Methods*

Six historical randomised control trials were selected for use in both the creation of the MAP prior and synthetic control arm. Using simulated trial data, we compared power, type 1 error, bias, coverage, and effective sample size. For the latter, SCM analyses were adjusted to match the total information contributed by the BDB prior plus concurrent trial participants.

*Results*

BDB MAP prior produced a mean response rate (historical control response probability) of 0.22 (posterior 95% credible interval: 0.11-0.36) (0.22 and 0.11-0.37 robustified), a power of 0.58, a type 1 error rate of 0.03, bias of 0.0007 (0.0008 robustified), a coverage of 0.96 (0.94 robustified) and prior ESS of 8.9 (3.32 robustified). SCM produced a mean response rate of 0.26 (95% confidence interval: 0.17-0.36), a power of 0.75, type 1 error rate of 0.013, bias of -0.0002, and coverage of 0.97; when adjusted to match the BDB ESS, SCM reported a similar mean response rate (0.27 and 0.13-0.46). a lower power (0.54), type 1 error rate (0.002) and bias(-0.0006), and the same coverage rate (0.97) to the initial SCM results.

*Discussion*

For this case study, the sample size adjusted SCM model produced a relatively similar power and type 1 error rate to the BDB MAP prior method. However, direct comparisons between these two methods fail to capture the true relationship due to how the data is handled differently and the purpose of each method. Method selection should consider trial-specific constraints, prior-data heterogeneity, and the desired balance between efficiency and robustness.

## Introduction

When a randomised controlled trial (RCT) is expected to struggle with recruitment or retention, novel methodologies can be proposed to augment or decrease the necessary sample size. Bayesian dynamic borrowing (BDB) and synthetic control methods (SCM) are both used in clinical trial design to avoid the need for prospective control arms (1,2). Rare diseases, pediatric studies, and trials where there is a large difference between the standard of care and the proof-of-concept drug effects are all examples where external control or data borrowing methodologies can improve efficiency or feasibility (3–5).

BDB and SCM address the challenge of overcoming the need for a prospective control arm differently. At the trial design stage, BDB creates a prior distribution of the control response rate informed by historical data, which carries information from the outcomes of participants from previous trials with similar populations and the same standard of care/placebo (6,7). This prior is combined with hypothetical new data through simulation to measure the contribution power of the prior. The posterior distributions from these simulations can be used to evaluate the operating characteristics of the design, including the power of the study at a pre-specified type 1 error rate, and to determine the number of control patients required to achieve that power. In dynamic borrowing, the influence of the prior is adjusted based on the similarity between the historical data and observed trial data. If the observed control response rate (or other outcome of interest) in the current study differs substantially from what is predicted by the historical prior — as detected through increased between-study heterogeneity — the model automatically downweights the prior's contribution to the posterior (8,9).

Synthetic control methods generate a control arm based on previous study data. These control arms have similar observed patient-level variables, outcome distributions, and response rates to the studies they are based on (3,10). Popular approaches include propensity score matching on baseline covariates or linear regression weighting, although decision trees like CART (classification and regression tree) can also be used (11–13). The generated data can be used to augment an existing control arm, increasing the overall sample size; this is referred to as a hybrid synthetic control (14). The generated data can also be used directly as a full synthetic control arm. In this case, a clinical trial would recruit only for the treatment arm and the findings would be compared to the synthetic control arm directly. This allows the trial to emulate the robustness of a RCT.

There are benefits and drawbacks to both methods. For example, while all methods can incorporate numerous historical studies, only BDB can directly account for the variability between these studies. In BDB, the contribution of historical data to the posterior distribution of the control response rate is weighted according to the consistency of the observed trial outcomes with the prior. Specifically, the robustified MAP prior downweights historical studies that are substantially different from the newly observed control responses, and this adjustment is quantified via the effective sample size of the prior.

On the other hand, SCM has a more “one and done” approach. The synthetic control arm is produced and then analysed as a control arm would typically be, providing a relatively simple approach. SCM can also incorporate covariates that may influence the outcome, such as sex or ethnicity, when generating the synthetic control. SCM may be particularly useful for studies where including a concurrent control arm presents challenges, such as when the treatment is expected to have a substantially improved effect over the standard of care. In contrast, BDB can incorporate historical information through a MAP prior, with the extent of borrowing determined according to the compatibility between the historical and current trial data. Therefore, in settings where the borrowing weight is informed by the similarity between historical and current data, some concurrent control-arm data may be required to assess this compatibility.

Both methods are highly dependent on the quality of the historical data available (5,10,15). If BDB is given data with high variability between studies and unclear patterns, a weak prior will be formed, and the analysis will gain less efficiency. If SCM are used with poorly fitting historical data, the synthetic control group may not be representative, resulting in misleading conclusions.

Directly comparing BDB and SCM is a challenge for several reasons. First, their ideal applications are different. BDB is typically employed when recruitment for the control arm is challenging, allowing historical control data to be borrowed to reduce the number of participants needed in that arm while still maintaining statistical power (16,17). The goal is to minimize the total number of participants in the control group while preserving rigorous trial evaluation. SCM, in contrast, is particularly suitable for trials where assigning participants to the control arm is difficult or ethically undesirable, such as when a highly effective treatment exists (10).

Second, what each method produces is inherently different. The product of BDB is a potentially informative prior that can be incorporated at the design stage to evaluate operating characteristics and inform the required size of the concurrent control group. In contrast, the product of SCM is the synthetic control group itself, which is constructed based on the characteristics of the intervention arm and historical data. This may make SCM more difficult to use for sample size determination at the design stage, as the quality and size of the synthetic control may depend on the availability and comparability of suitable historical data and covariates. However, when the BDB borrowing weight is determined according to the compatibility between historical and current trial data, the extent of borrowing is also not fully known at the design stage.

The assumptions and data needs of each method also differs. BDB can leverage aggregate-level data from historical trials, whereas SCM generally requires individual-level data to synthesize a control arm. BDB is applied at the trial design stage as it impacts the size of the control arm needed for recruitment. SCM can also be considered at the design stage, where historical data are used to construct a synthetic control in place of a conventionally recruited control arm. SCM can also be applied post-hoc for single-arm studies looking to compare against a hypothetical control arm to mimic randomisation, although this does not provide the randomisation of a conventional RCT.

The performance of both methods can be evaluated using common statistical measures. For BDB, frequentist operating characteristics such as power, type 1 error rate, bias and coverage can be assessed through repeated sampling or simulation to evaluate the performance of the Bayesian design (16,18). These measures can also be applied to SCM to assess the performance of the synthetic control under repeated simulated trials. In addition, SCM requires assessment of how well the synthetic control represents the treatment arm, which can be evaluated by comparing covariate and outcome characteristics between the synthetic control and treatment groups (10,19).

In addition to BDB and SCM, other approaches exist for incorporating external control data into clinical trial design. For example, two-stage designs and propensity score–integrated borrowing methods combine external and trial data through weighting or adjustment techniques, often using covariate information to improve comparability between historical and current populations (11,20–23). These approaches lie between fully model-based borrowing, such as BDB, and fully constructed control arms, such as SCM. There are also a number of variations of Bayesian and synthetic methods aside from the ones used in this study, including hybrid synthetic control approaches where real data is combined with synthetic data (14,24–26). It is possible to combine Bayesian and synthetic methods in a single study, using Bayesian methods for trial monitoring or decision-making and synthetic control methods to augment or replace control data (27). This study focuses on BDB and SCM as two widely used and conceptually distinct approaches to incorporating external data, representing model-based borrowing and data-driven control construction, respectively.

BDB and SCM have both been applied in clinical trial settings, but their performance may differ depending on the context. To address this, they are compared within a unified case study of pediatric atopic dermatitis using historical randomised controlled trials.

Atopic dermatitis is the most common chronic inflammatory skin condition (28)Recent clinical advancements have identified more targeted treatments, such as topical PDE4 inhibitors, oral JAK inhibitors, and injected biologics, such as dupilumab, tralokinumab, lebrikizumab, and others (29,30); these treatments have shown much more effective treatment in adults over the past decade. Atopic dermatitis affects 5-20% of children (31), while affecting adults significantly less (prevalence of adult atopic dermatitis varies heavily based on region). Pediatric cases are also more likely to be severe than adult-onset cases (32–35). The need for effective clinical testing of new treatments for atopic dermatitis on a pediatric population is evident.

Clinicians often struggle to recruit for pediatric RCTs. Pediatric RCTs are more susceptible to trial discontinuation and nonpublication due to challenges with enrolling viable sample sizes (36). Additional ethical considerations are required by regulators in order to perform pediatric RCTs, making them more challenging to design (37). Involvement and reliance on caretakers for recruitment and adherence of the patient is another complexity of pediatric trials, leading to lower rate of recruitment and lower rates of adherence during trials (38,39).

Pediatric atopic dermatitis was chosen as a case study due to the necessity of running clinical trials on developing treatments and the difficulty in recruitment that SCM and BDB may be able to address.

Performance was evaluated using power, type 1 error rate, bias, and coverage. This paper attempts to outline a method of comparing BDB and SCM using these metrics, but it is important to acknowledge the limitations of trying to compare methods with fundamentally different inputs, outcomes, and typical metrics. The reason behind the comparison is that their practical applications to clinical trials aim to solve the same problems, and therefore they are often discussed as main alternatives to each other. As of this paper, there is no current way of deciding which method is best based on statistical metrics. This study is an exploration of whether that is possible.

**Methods**

Six historical RCTs were used in both the BDB and the SCM models (30,40–44). Placebo group sample sizes and response rates were included from each of the six historical RCTs. The response rate refers to the number of patients that reached the desired outcome of EASI-75 (75% improvement in the Eczema Area and Severity Index). A full list of the included studies can be found in Table 1.The R library RBesT (45) was used to create a MAP prior for the BDB method. For the MAP prior, historical study control responses were modelled with a binomial likelihood and a hierarchical random-effects structure. The resulting MAP prior was approximated as a mixture of beta distributions. The prior was robustified by adding a vague component with a specified weight (50% in this case), allowing the analysis to downweight historical studies that differ from the trial data. The effective sample size of the prior was assessed to quantify its influence on the operating characteristics.

| **Study** | **Sample Size (responders/total control group size)** | **Intervention** | **Outcome** | **Participant Age Range (years)** |
|---|---|---|---|---|
| Efficacy and Safety of Dupilumab in Adolescents with Uncontrolled Moderate to Severe Atopic Dermatitis: A Phase 3 Randomized Clinical Trial *Simpson, E.L. et al. (2020)* | 7/85 | Dupilumab monotherapy | EASI-75 | 13-17 |
| Efficacy and safety of dupilumab with concomitant topical corticosteroids in children 6 to 11 years old with severe atopic dermatitis: A randomized, double-blinded, placebo-controlled phase 3 trial *Paller, A.S. et al. (2020)* | 33/123 | Dupilumab + topical corticosteroids | EASI-75 | 6-11 |
| Dupilumab in children aged 6 months to younger than 6 years with uncontrolled atopic dermatitis: a randomised, double-blind, placebo-controlled, phase 3 trial *Paller, A.S. et al. (2022)* | 28/79 | Dupilumab + topical corticosteroids | EASI-75 | 6 months – 6 years |
| Efficacy and safety of baricitinib in combination with topical corticosteroids in paediatric patients with moderate-to-severe atopic dermatitis with an inadequate response to topical corticosteroids: results from a phase III, randomized, double-blind, placebo-controlled study (BREEZE-AD PEDS) *Torrelo, A. et al. (2023)* | 39/122 | Baricitinib + topical corticosteroids | EASI-75 | 2-18 |
| Efficacy and Safety of Tralokinumab in Adolescents with Moderate to Severe Atopic Dermatitis: The Phase 3 ECZTRA 6 Randomized Clinical Trial *Paller, A.S. et al. (2023)* | 6/94 | Tralokinumab monotherapy | EASI-75 | 12-17 |
| Efficacy and safety of dupilumab with concomitant topical corticosteroids in Japanese pediatric patients with moderate-to-severe atopic dermatitis: A randomized, double-blind, placebo-controlled phase 3 study *Ebisawa, M. et al. (2024)* | 6/32 | Dupilumab + topical corticosteroids | EASI-75 | 6-12 |

Table 1: **Included Study Data.** Each of the six studies used in both methods is listed above, including their control group sample sizes, interventions, outcome, and participant age range.

*Hypothesis*

The planned sample size for the new treatment group is 30 patients. This value is consistent across both the BDB and SCM frameworks.

The expected treatment effect for the new treatment group is an increase of 0.24 in the probability of achieving EASI-75 compared to the control group. This effect was chosen based on clinical considerations and is applied consistently in both BDB and SCM frameworks.

The parameter of interest, θ, is the difference in the probability of achieving EASI-75 between a hypothetical treatment arm and the control arm:

$$\theta = p_{\text{treatment}} - p_{\text{control}}$$

The control response rate is derived either from the MAP prior (BDB) or from synthetic data (SCM), and a treatment effect of 0.24 is applied to define the assumed treatment response rate.

The one-tailed hypotheses can be stated as follows:

$$H_0: \theta \leq 0$$

$$H_1: \theta > 0$$

*SCM*

The R library Synthpop (13) was used to create a synthetic control arm for the SCM. Historical control outcomes from the six included trials were simulated and combined into a single dataset. For each participant, a binary response $y_{ij}$ was defined, with $j = 1, \ldots, n_i$ within study $i$. The pooled control response probability was calculated as:

$$\hat{p}_{\text{control}} = \frac{\sum_{i=1}^{m} \sum_{j=1}^{n_i} y_{ij}}{\sum_{i=1}^{m} n_i}$$

Where $m$ is the number of historical studies (in this case, 6).

CART was applied to model the distribution of the binary outcome, resampling responses within terminal nodes to preserve the empirical distribution of historical control data. In each simulation replicate, a new synthetic dataset was created to represent a plausible control arm for a future trial, and the number of responders was drawn according to this underlying probability. Each synthetic dataset contained participant counts equal to the mean historical sample size.

For synthetic control data generated with CART, the following notation shows the synthesis process:

Let $T$ represent the tree structure.

$T$ is composed of nodes $N$ and edges $E$ such that $T = (N, E)$ .

At each node $N_i$ a feature $X_j$ and a threshold $t$ are selected to split the data.

The splitting criterion can be written as:

$$split(N_i) = \{X_j < t\} \; or \; \{X_j \geq t\}$$

This produces two child nodes $N_{i\,L}$and $N_{i\,R}$.

The data $D$ is recursively partitioned based on the splitting criteria:

$$D = D_L \cup D_R \text{ where } D_L = \{X_j < t\}, D_R = \{X_j \geq t\}$$

This process is continued until the stopping criterion, either the minimum node size or the maximum tree depth, is met.

For a leaf node $L_k$, the prediction $\hat{y}_k$ for a new data point $x$ is:

$$\hat{y}_k = \frac{1}{|L_k|} \sum_{(x_i, y_i) \in L_k} y_i$$

Synthetic responses $\tilde{y}$ are generated:

$$\tilde{y} \sim N(\hat{y}k, \sigma_k^2)$$

Where $\sigma_k^2$ represents the variance of the responses within $L_k$.

For any new data point $x$, the tree $T$ is followed using the splitting criteria until a new leaf node $L_k$is reached. A new synthetic response $\tilde{y}k$ is then generated.

*BDB*

Historical control response data from six randomised trials were incorporated using a hierarchical meta-analytic model.

For study $j$, the number of responders $r_j$among $n_j$ control patients were assumed to follow:

$$r_j \sim \text{Binomial}(n_j, p_j)$$

with study-specific response probabilities (for between-study heterogeneity) modelled on the logit scale as:

$$\text{logit}(p_j) = \mu + u_j$$

where

$$u_j \sim N(0, \tau^2)$$

The between-study heterogeneity parameter was then assigned a half-normal prior:

$$\tau \sim \text{HalfNormal}(0,1)$$

The hierarchical model was fitted using the gMAP implementation in the RBesT package. The resulting meta-analytic predictive (MAP) distribution for the control response probability was represented by a mixture of up to three beta distributions using automixfit(Nc=1:3). To protect against prior-data conflict, the MAP prior was robustified by assigning 50% prior weight to the MAP distribution and 50% to a vague component. The resulting robustified prior was used for the placebo/control arm of the Bayesian dynamic borrowing analysis. The treatment response probability was assigned a Beta(1,1) prior.

The fitted hierarchical model generates the MAP distribution:

$$p_C \mid D_h \sim \text{MAP prior,}$$

where $D_h$represents the six historical control datasets.

The MAP distribution is approximated by a beta mixture with up to three components:

$$p_C \sim \sum_{k=1}^{K} w_k \,\text{Beta}(a_k, b_k), K \leq 3.$$

The MAP prior for the new trial control arm is obtained by integrating over the posterior predictive distribution of $\theta$, where $\theta$ represents the control response rate (the parameter of interest) in the new study.

Let $D_h = \{y_1, \dots, y_m\}$ represent the historical control data from $m$ previous trials.

Each historical study $i$ has sample size $n_i$ and observed responses $y_i$.

$$p(\theta \mid D_h) = \int \prod_{i=1}^{m} \text{Binomial}(y_i \mid n_{i,} \theta_i) \cdot p(\theta_i \mid \alpha, \beta)\, d\theta_i$$

To increase robustness to potential conflicts between historical and new data, the MAP prior is robustified by adding a vague component:

$$P_{\text{robust}}(\theta) = w \cdot p_{\text{MAP}}(\theta) + (1 - w) \cdot \text{Beta}(1,1), w \in [0,1]$$

Let $D_n = \{y_{new}\}$ be the planned data from the new control arm.

The posterior for $\theta$ is:

$$p(\theta \mid D_h, D_n) \propto p(D_n \mid \theta) \cdot P_{\text{robust}}(\theta)$$

Where $P_{\text{robust}}(\theta)$ is the robust MAP prior.

Effective sample size (ESS) is reported for the BDB method, but not for SCM as a synthetic control arm would fully replace a recruited control arm and can be generated at any sample size to match the treatment arm. In order to provide a degree of comparison, an additional computation of the SCM was performed with the produced sample size being set to equal the ESS of the robust prior of the BDB method so that metrics can be compared with the same sample size assumption.

ESS in BDB is achieved through the following:

$$\mathrm{ESS} = \mathrm{ESS}_{\mathrm{prior}} + n_{\mathrm{control}}$$

Where $\mathrm{ESS}_{\mathrm{prior}}$ refers to the weight the prior carries in the sample size. This is added to the sample size of the control group used in the calculation (n=30) to get the total ESS.

The operating characteristics of the Bayesian design were evaluated using the oc2S() function in RBesT, assuming 30 patients in each treatment arm. The Bayesian decision criterion was defined as a posterior probability greater than 0.975 that the treatment response probability exceeded the control response probability. Power was calculated under scenarios in which the treatment response probability exceeded the control response probability by 0.24, while the type I error rate was calculated under scenarios in which the treatment and control response probabilities were equal. The oc2S() function evaluates the probability of meeting the decision criterion over the possible outcomes of the planned trial under each assumed pair of true response probabilities. To evaluate posterior estimation performance, 1,000 simulated new control-arm datasets of 30 patients were generated assuming a prespecified true control response probability. For each simulated dataset, the historical MAP and robustified MAP priors were updated using postmix() to obtain the corresponding posterior distributions. Posterior means and 95% posterior credible intervals were obtained for each simulation. Bias was calculated as the posterior mean minus the true control response probability, and coverage was calculated as the proportion of simulations in which the 95% posterior credible interval contained the true control response probability. Mean credible-interval width was also recorded.

Although BDB uses a Bayesian posterior probability as its decision criterion, its operating characteristics were evaluated under specified frequentist data-generating scenarios. The BDB decision criterion was defined as:

$$P(\theta > 0 \mid \mathrm{data}) \geq 0.95$$

corresponding to a conclusion that the treatment has a positive effect.

The false-positive rate was evaluated under the scenario in which the true treatment effect was zero, with the active and placebo response probabilities set equal to the mean of the MAP prior. The probability of a true-positive conclusion (referred to as power) was evaluated under a treatment-effect scenario of $\theta = 0.24$, with the active response probability set to the control response probability plus 0.24. Therefore, the reported type I error and power represent frequentist operating characteristics of the prespecified Bayesian decision rule.

To assess the sensitivity of the BDB operating characteristics to the assumed control response rate, the control-arm EASI-75 response probability was varied across a clinically plausible range informed by the historical trials. The observed control response rate in each study was calculated as the number of EASI-75 responders divided by the control-arm sample size. Across the six studies included in the MAP prior, observed control response rates ranged from 6.4% to 35.4%, with the average being 0.24.

Accordingly, the BDB operating characteristics were evaluated across this range of control response probabilities. For the power analysis, the treatment response probability was defined as:

$$P_{treatment} = P_{control} + 0.24$$

maintaining the assumed treatment effect of 0.24 across the range of control response rates.

For the false-positive rate analysis, the treatment and control response probabilities were set equal, corresponding to a true treatment effect of $\theta = 0$. This sensitivity analysis was used to assess how the probability of meeting the prespecified Bayesian decision criterion varied with the underlying control response rate.

*Comparison*

The R packages were chosen for their transparency, reproducibility, and established use in applied research (8,45–47). Both approaches rely on assumptions regarding historical data: the robustified MAP prior assumes exchangeability but allows downweighting of conflicting studies, and the synthetic control approach assumes historical relationships adequately represent potential trial outcomes. Figure 1 shows a flowchart of steps for both methods. The full R code for both methods is available on GitHub via the following link: [https://github.com/N-cizauskas/Bayesian-vs-Synthetic](https://github.com/N-cizauskas/Bayesian-vs-Synthetic)

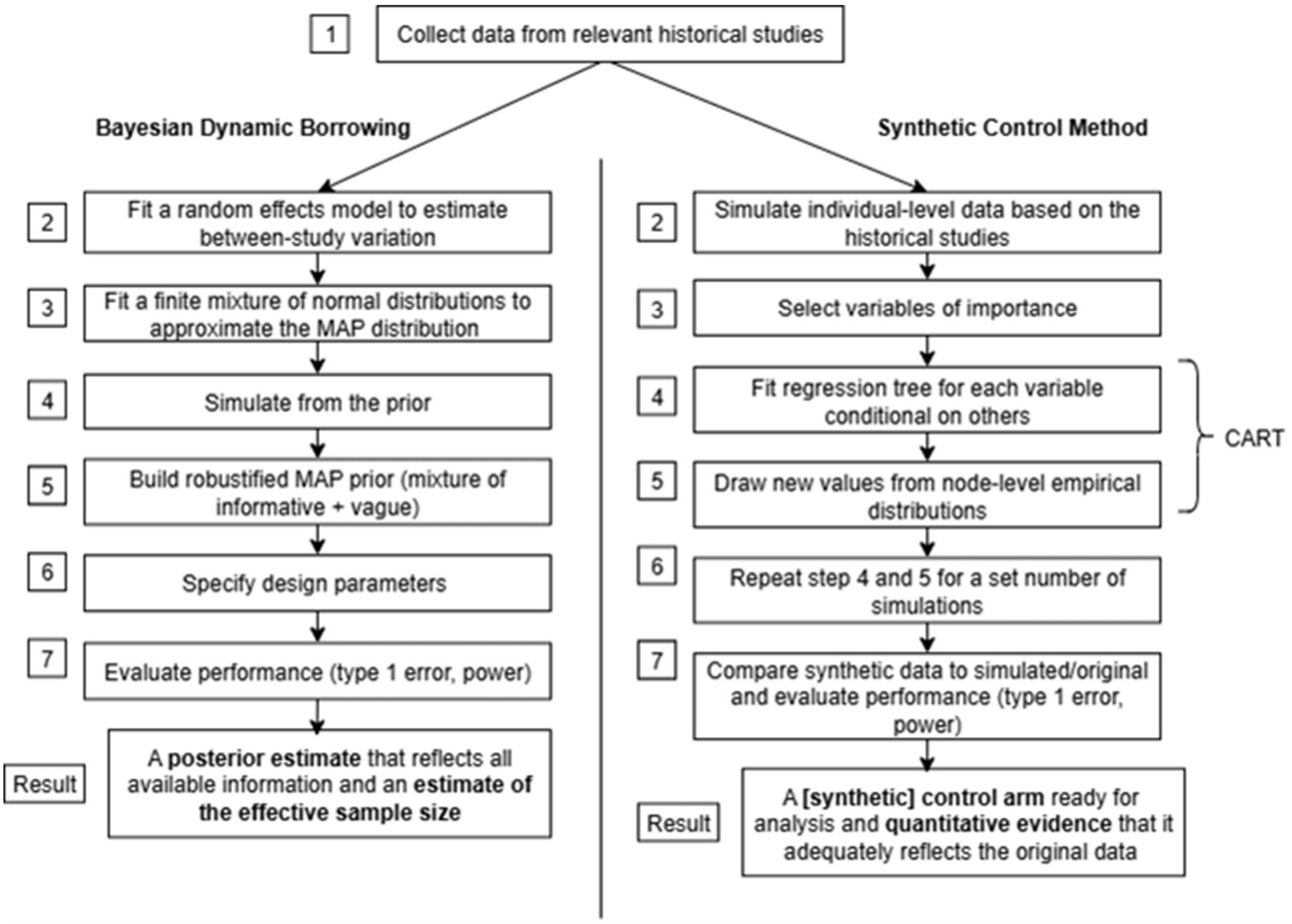

Figure 1: **Flowchart of Steps for BDB and SCM.** Steps to perform and apply BDM and SCM starting with relevant historical studies and ending with the result of each method.

The target population includes children with atopic dermatitis, consistent with the populations represented in the historical trials.

Comparison metrics included mean response rate, confidence/credible intervals type 1 error rate, power, bias, and coverage.

Although both confidence and credible intervals quantify uncertainty, they arise from different statistical frameworks. In large datasets, credible and confidence intervals often align closely because the likelihood dominates the prior, whereas in smaller datasets an informative prior—as in Bayesian dynamic borrowing—can yield narrower credible intervals (48). It is important to note that this is not a frequentist measurement of repeated measures coverage, but a Bayesian equivalent.

For BDB, bias was measured as the difference between the posterior mean and the true placebo rate (mean of MAP prior for BDB). For SCM, bias is the difference between the synthetic control estimated response rate and the reference historical control response rate.

Across 10,000 simulated replicates, power and type 1 error were estimated as the proportion of simulations in which the prespecified decision criterion was met. For the SCM, the treatment effect was assessed using Fisher's exact test, while the BDB analysis used the prespecified Bayesian posterior probability decision criterion.

For both BDB and SCM, power was evaluated at a common nominal Type I error rate of $\alpha = 0.05$

Table 2 contains simulation parameters for both BDB and SCM.

| Method | Control Response Probability | Treatment Effect | Sample Size (control arm) | Number of Simulations | Other Settings |
|---|---|---|---|---|---|
| BDB | (MAP prior mean) 0.243 | 0.24 | 30 | 10,000 posterior draws per trial | Hierarchical random effects, robustified prior weight of 0.5, binomial likelihood |
| SCM | (pooled historical mean) 0.222 | 0.24 | 30 | 10,000 | CART method |

Table 2: **Simulation Parameters.** Parameters of both BDB and SCM are listed for the purpose of potential replication.

*Covariates*

Covariates were not included in either the main BDB or SCM implementations. For the BDB approach, only study-level control arm outcomes (number of responders and sample size) were used to fit the robustified MAP prior, which accounts for between-study heterogeneity but not individual patient characteristics. For the SCM approach, Synthpop generated synthetic control arms using only the observed response counts from the historical studies; CART was used to model distributions and resample responses and the trees operated solely on the binary outcome. The purpose of excluding covariates is to keep the methods as comparable as possible; the RBest package does not easily implement covariates into the BDB method, so covariates were also excluded from the SCM method. However, one of the main benefits of SCM, CART especially, is the ability to recreate covariate relationships. To capture this, two sensitivity analyses including covariates in only the SCM method were included.

The first sensitivity analysis included two covariates available in all historical studies, sex and age.

An important note is that the data for this study was taken from publicly available summary-level statistics from the listed studies, and the covariate analysis therefore captures the marginal distributions but may not reflect the relationships between the covariates and the responses. Another challenge of using summary-level statistics was the difference in how covariates were reported. Almost every study reported age as a categorical variable using a different grouping (e.g. age under 2 versus over 2 or age under 12 versus over 12); as a result, the covariate values for age were simplified.

The secondary sensitivity analysis of the SCM was performed using hypothetical covariates with inferred relationships. This model is not based on real data from the studies. The purpose of including this is to show how the values would hypothetically change if the covariates had relationship level information.

The relationship was specified using a logistic model:

$$logit\{P(Y = 1)\} = \alpha + \beta_1 X_1 + B_2 X_2$$

Where $Y = 1$ represents response, $X_1$ represents female sex, and $X_2$ represents age under 2 years. The regression coefficients are the natural logarithms of the odds ratios:

$$\beta_1 = \log(1.5) = 0.405$$

and

$$\beta_2 = \log(2) = 0.693$$

The model therefore gives each participant a relative probability of response based on their sex and age. A male participant aged over two years is the reference group, with a relative weight of 1. A female participant aged over two years has an odds ratio of 1.5, a male participant under two has an odds ratio of 2, and a female participant under two has an odds ratio of 1.5 x 2 = 3. The intercept was set to $\alpha = 0.5$. This simulated data contains a known relationship in which female sex and younger age are associated with greater response.

The number of responders in each study was then kept equal to the observed number of responders. Participants were selected as responders according to their relative weights, meaning that participants with higher weights were more likely to be assigned a response.

The resulting data were therefore used as an "ideal" scenario in which the relationship between covariates and response was known in advance, allowing us to assess whether the synthetic data-generation method was able to reproduce that relationship. The values used in this scenario are not reflective of the real data or findings from the studies and are only used to showcase the methodology.

## Results

*BDB*

The MAP prior method had a mean of 0.24 (95% credible interval: 0.02-0.69), and the robustified MAP prior method had a mean of 0.37 (95% credible interval: 0.03-0.95). The MAP posterior mean was 0.22 (95% credible interval: 0.11-0.36) with a bias of 0.0007, and the robustified MAP posterior mean was 0.22 (95% credible interval: 0.11-0.37) with a bias of 0.0008.

The MAP prior method had a power of 0.58 and a type 1 error rate of 0.03. The robustified MAP prior method had a power of 0.58, a type 1 error of 0.03. The MAP prior had an effective sample size of 8.92, which decreased to 3.32 after robustification.

The effective sample size of the prior was 8.92, and when combined with the active and control arms of 30 patients each (with a prior weight of 0.5), this corresponds to a control arm size of approximately 38.92 patients. The robustified prior was 3.32, which corresponds to approximately 33.32 patients.

The range for power was from 0.517 to 0.693, and the range for type 1 error rate was from 0.006 to 0.039. The power and type 1 error curve graphs from the Bayesian sensitivity analysis can be seen in Figure 2.

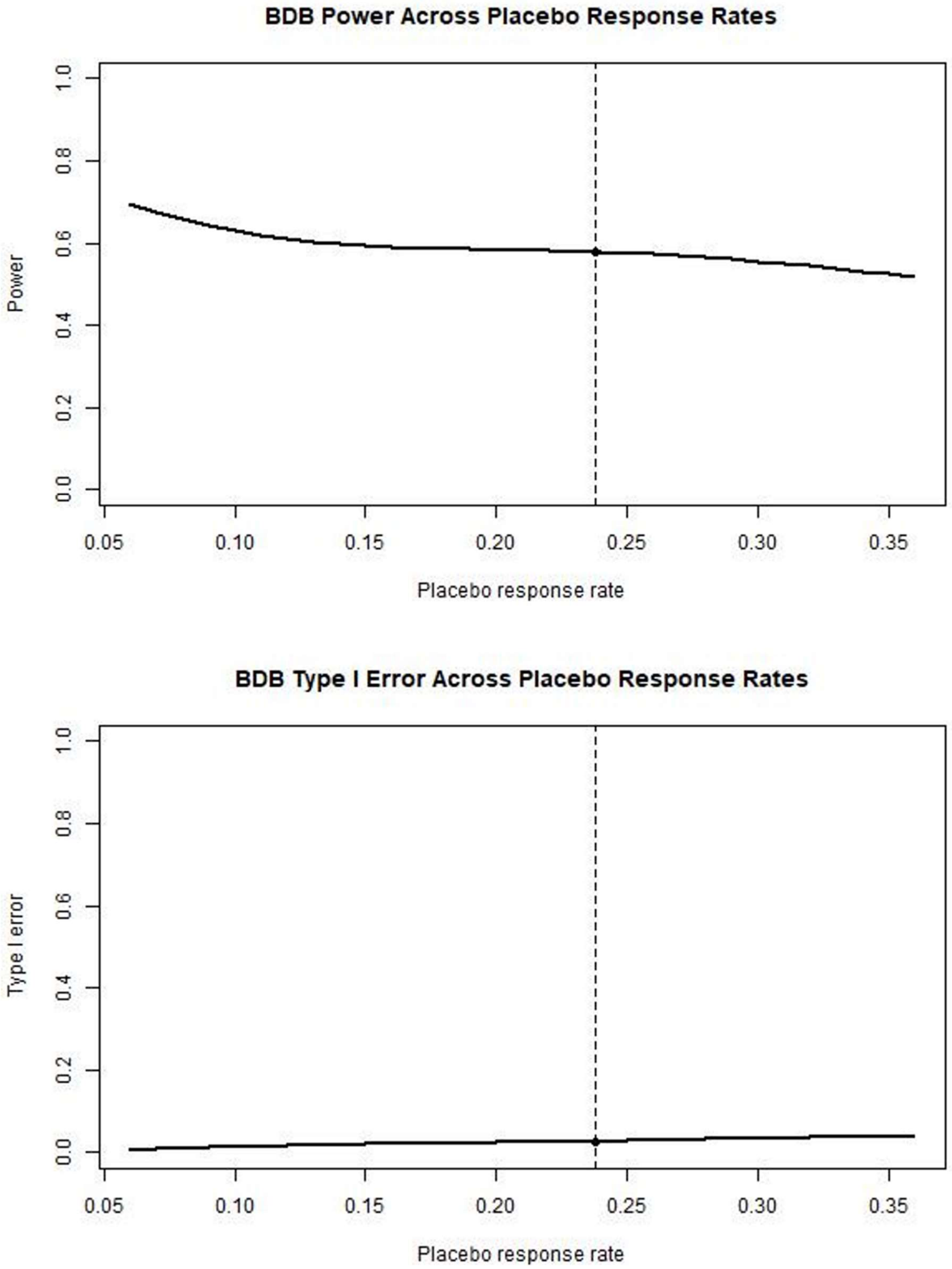

Figure 2: **Power and Type 1 Error Curves** The top line plot shows the power of the MAP prior across a set of potential placebo response rates as a comparison to the assumed treatment effect of 0.24. The bottom line plot shows the type 1 error of the MAP prior across the same set of potential placebo response rates.

While the power and type 1 error being the same between MAP prior and robustified MAP prior being the same may be surprising, this is only due to rounding. The true values differed by very small amounts, therefore rounding to the same value. The mean response rates for the prior and robustified prior differ by 0.13, but the mean response rates for the posterior and robustified posterior only differ by 0.02.

*SCM*

SCM produced a mean response rate of 0.26 (95% confidence interval: 0.17–0.36), a power of 0.75, a type 1 error rate of 0.013, a bias of -0.0002, and a coverage rate of 0.97.

The SCM model synthesized a control group based on the mean sample size of the historical studies (n = 89.2). The difference in model sample size reflects the design philosophies of the two methods: BDB aims to minimize new data collection by borrowing information from the prior, whereas SCM can generate any synthetic control size based on the available historical data, so minimization is not its primary goal.

To enable a direct comparison, we adjusted the SCM analysis to use a control group of n = 33 (approximately the same as the BDB total). Under this scenario, the SCM achieved a mean response rate of 0.27 (95% confidence interval: 0.13-0.46) power of 0.54, type 1 error rate of 0.002, bias of -0.0006, and coverage rate of 0.97.

SCM with covariates produced a mean response rate of 0.21 (95% confidence interval: 0.13–0.31), a power of 0.75, a type 1 error rate of 0.014, a bias of 0.0002, and a coverage rate of 0.97.

SCM with covariates adjusted to control group n = 33 produced a mean response rate of 0.18 (95% confidence interval: 0.07–0.35), a power of 0.54, a type 1 error rate of 0.002, a bias of 0.0009, and a coverage rate of 0.97.

SCM with ideal covariates produced a mean response rate of 0.22 (95% confidence interval: 0.14–0.33), a power of 0.75, a type 1 error rate of 0.013, a bias of -0.0002, and a coverage rate of 0.97.

SCM with ideal covariates adjusted to control group n = 33 produced a mean response rate of 0.21 (95% confidence interval: 0.09–0.39), a power of 0.54, a type 1 error rate of 0.002, a bias of 0.0008, and a coverage rate of 0.97.

While the coverage being the same for all SCM sensitivity tests being the same may appear to be alarming, this is only due to rounding. The true values differed by very small amounts, therefore rounding to the same value.

*Comparison*

The response rate in the control arm of both the BDB and SCM were based on the combined response rates of historical studies. In Figure 3, the response rates of the historical studies (including their mean) and of BDB and SCM controls are shown in a forest plot.

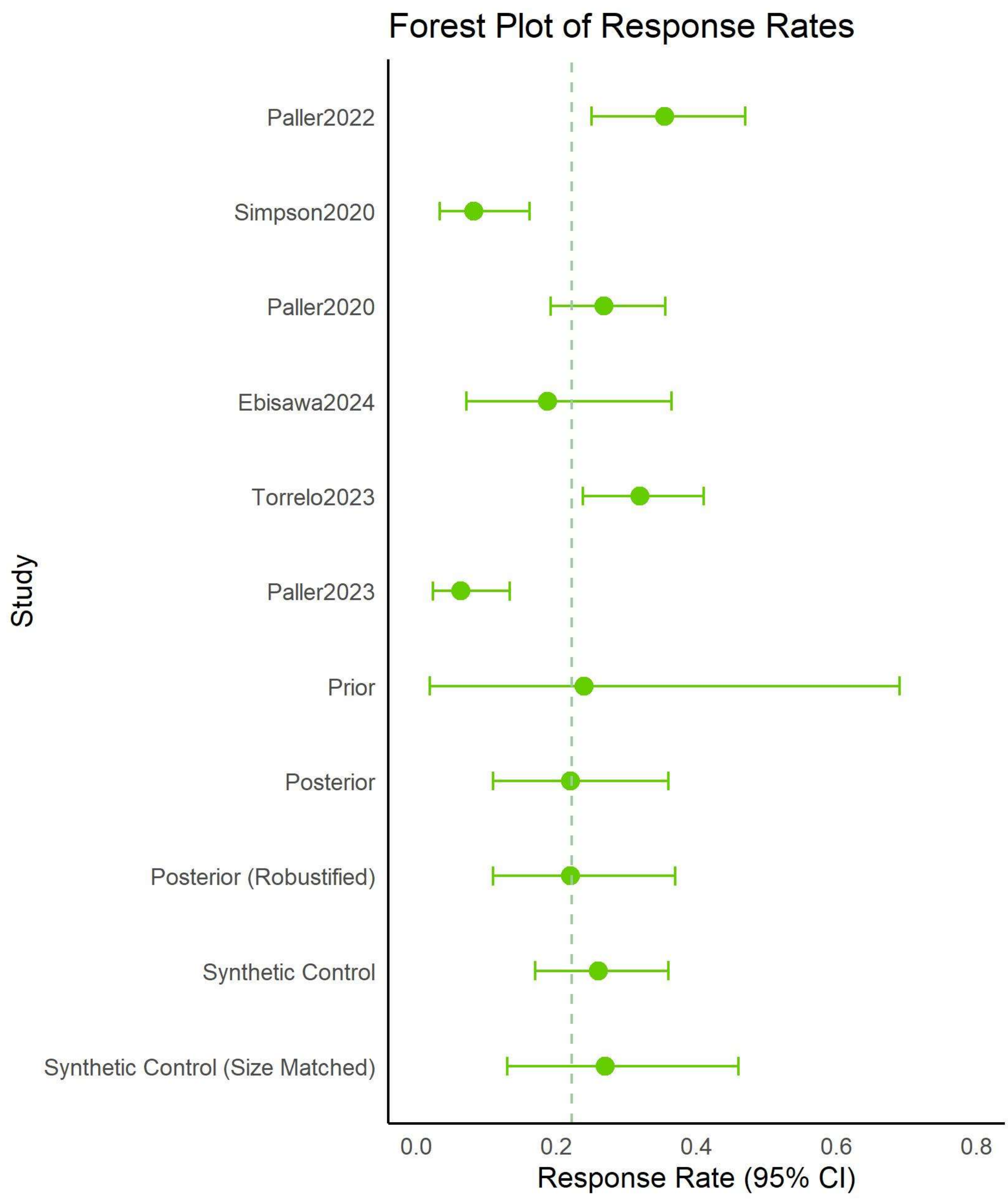

Figure 3: **BDB and SCM Forest Plot.** The forest plot compares the response rate and 95% confidence/credible interval for historical studies, combined mean, MAP prior, MAP posterior, and synthetic control.

All results are outlined in Table 3 for comparison.

| Method | Mean Response Rate (Confidence or Credible Interval at 95%) | Power | Type 1 Error | Bias | Coverage | ESS |
|---|---|---|---|---|---|---|
| SCM (n=89) | 0.26 (0.17–0.36) | 0.75 | 0.013 | -0.0002 | 0.97 | -- |
| SCM (n=33) | 0.27 (0.13-0.46) | 0.54 | 0.002 | -0.0006 | 0.97 | -- |
| SCM with Covariates (n=89) | 0.21 (0.13–0.31) | 0.75 | 0.014 | 0.0002 | 0.97 | -- |
| SCM with Covariates (n=33) | 0.18 (0.07–0.35) | 0.54 | 0.002 | 0.0009 | 0.97 | -- |
| SCM with Ideal Covariates (n=89) | 0.22 (0.14–0.33) | 0.75 | 0.013 | -0.0002 | 0.97 | -- |
| SCM with Ideal Covariates (n=33) | 0.21 (0.09–0.39) | 0.54 | 0.002 | 0.0008 | 0.97 | -- |
| BDB MAP Prior | 0.24 (0.02-0.69) | 0.58 | 0.03 | -- | -- | 8.92 |
| BDB MAP Posterior | 0.22 (0.11-0.36) | -- | -- | 0.0007 | 0.96 | -- |
| BDB Robustified MAP Prior | 0.37 (0.03-0.95) | 0.58 | 0.03 | - | -- | 3.32 |
| BDB Robustified MAP Posterior | 0.22 (0.11-0.37) | -- | -- | 0.0008 | 0.94 | -- |

Table 3: **Results by Method.** Includes all primary evaluation metrics for both SCM, BDB, and all sensitivity analyses performed. Metrics that could not be calculated due to methodological incompatibility are marked with --.

**Discussion**

Overall, both approaches yielded comparable control response estimates (difference ≈ 5%), with highly overlapping uncertainty intervals – with one exception. The BDB robustified MAP prior had a higher mean response rate (0.37) than all other methods, and a much wider credible interval (0.03-0.95).; however, the BDB robustified MAP posterior had an average mean response rate (0.22) and a much narrower credible interval (0.11-0.37). The prior operating characteristics are based only on the historical study data and not on hypothetical new trial data, and as a result show very different results than all other methods.

There was a sizable difference between the power of SCM methods of differing sample sizes, with the larger sample size (n=89) consistently having a power of 0.75 and the smaller sample size (n=33) consistently having a power of 0.54. This is expected, as larger data sets generally increase statistical power (49).

It is not straightforward to compare BDB and SCM models due to the inherent differences in quality metrics for Bayesian and frequentist approaches – for example, the results showed a comparison between the credible interval and confidence interval, but these are not the same metric and should not be directly compared as such. Furthermore, BDB and SCM have different applications in real world cases. BDB is used to determine a maximum sample size needed for a control group, and SCM is used to generate the control group itself. Practically, determining which method is "better" will depend on the specific needs of the study, such as how difficult recruitment is. In cases where recruitment for the control group is near impossible, SCM is a much better choice. SCM can generate the entire control group from historical data without any need for supplementation with real participants. In cases where recruitment as a whole is a challenge but allocation to the control group is not an issue, BDB may have more regulatory appeal.

The methodologies also incorporate the historical information differently: BDB is looking at the cumulative information from the intervals of previous studies, while SCM is creating a new interval based on the historical information.

The 95% credible interval reported for the MAP posterior (0.11-0.36) represents the posterior uncertainty in the parameter of interest. The 95% confidence interval reported for the synthetic control (0.17–0.36) summarises the uncertainty in the estimated mean difference between the synthetic control and the observed data under a frequentist framework.

While the credible interval and confidence interval summarise different types of uncertainty and are not directly comparable, this reflects the greater uncertainty in the BDB prior, which incorporated information directly and only from heterogeneous historical studies. In contrast, the synthetic control and the MAP posterior was generated directly from simulated individual-level data from the historical studies, leading to a narrower, more data-driven interval.

Both methods can also be used on the intervention arm as well. The purpose of this study is specifically comparing control groups, and therefore the results cannot be extrapolated to cases where an intervention arm is being augmented. The intervention effect may interact with trial-specific factors (e.g., population, noncommitment, dropout), making historical borrowing or synthesis less reliable (18,50,51). Methodologically, the prior or synthetic control would inform treatment effect estimates rather than baseline rates, introducing different risks and dependence on exchangeability assumptions. A future study would need to focus on comparing BDB and SCM in intervention arms with these differences in mind.

In addition to this, individual-level covariates were not initially incorporated into either the BDB or SCM analyses due to lack of consistent availability in the historical trials' summary data. The absence of covariate information means that potential differences in patient populations across the historical studies were not explicitly adjusted for. In practice, if substantial variation existed in important covariates across historical studies, both the MAP prior and synthetic controls could either over- or understate the effective information contributed by prior data.

RBest, the R package used for the BDB, does not have a function for including covariates. A covariate inclusion SCM was attempted, but was challenging due to the aforementioned lack of consistent availability of data. All studies reported age and sex, so these were selected to be test covariates. However, each study reported age as a different categorical grouping with little overlap. The included age covariate was heavily simplified. Another methodological issue in the inclusion of covariates was that the summary level statistics in each study reported the demographic information as overall measurements for the placebo group (e.g. the placebo group had x female patients and x male patients), instead of the preferred distribution of demographic information by response result within the placebo group (e.g. responders in the placebo group included x female and x male patients, non-responders in the placebo group included x female and x male patients). Because of this, only the marginal distributions of the covariates can be replicated during the CART process. The influence of covariates on the response is likely not captured adequately in these results. In order to adequately capture these relationships, patient-level data would be necessary.

The results of the covariate inclusive SCM showed very limited differences from the covariate exclusive SCM, further solidifying this point. It's important to note that we are not testing the effects of the covariates on the treatment effect here, but rather ensuring that the model (CART), which performs best when given covariates, was performing well.

Despite this limitation, the historical trials included in this analysis were relatively homogeneous pediatric populations, with similar inclusion criteria and outcome definitions (EASI-75). Therefore, the comparison between BDB and SCM in this case study is likely robust to moderate covariate variation. Future studies could extend these methods by incorporating patient-level covariates in the comparison.

An “ideal” scenario of SCM was produced using hypothetical relationships between the covariates and response (not based on the historical trials). This scenario aimed to capture how the methodology would be affected if the patient-level data were known. However, there was again very limited differences between the covariate exclusive SCM, the real covariate inclusive SCM, and the hypothetical ideal covariate SCM. From these two additionally scenarios, we can compare the results of the SCM to the BDB as confidently as possible given the restrictions of summary-level data.

The sensitivity analysis for BDB demonstrated that the evaluation metrics were relatively stable across the range of control response rates observed in previous paediatric atopic dermatitis trials. Despite variation in the assumed control response probability, both the probability of a positive treatment conclusion under the assumed treatment effect of 0.24 and the false-positive rate under a null treatment effect remained broadly unchanged. This suggests that, within the range examined, the reported BDB operating characteristics were not strongly dependent on the specific control response rate selected from the MAP prior.

Although the current BDB approach borrows information from all six historical trials simultaneously, performance may be sensitive to heterogeneity across these studies. An alternative strategy could involve a two-stage or hybrid borrowing approach, where study-level variability is explicitly accounted for before combining data into a single prior (6,21,52). Such methods could potentially increase the effective sample size contributed by historical data while reducing the risk of over-weighting inconsistent trials. While we did not implement this approach in the present analysis, acknowledging its existence highlights a potential avenue to improve the robustness of BDB in contexts with heterogeneous historical evidence.

Similarly, there are models for generating synthetic controls that may out-perform CART in power and error reduction, such as Generative Adversarial Networks (53,54). The aim of this study was to compare a Bayesian method to a frequentist synthetic control method, and two simple model designs were chosen to represent each. A future study could compare more alternative methods within and between Bayesian and synthetic methods.

The results indicated that both methods are viable and similarly powerful approaches for incorporating historical data. SCM with n=89 performed the best, with the highest power (0.75) and a relatively small type 1 error rate (0.013). The power between SCM at n=33 and BDB methods was similar (0.54 and 0.58 respectively) but the type 1 error was typically smaller in SCM methods (0.002 to 0.03 respectively).

The bias was extremely low, likely due to the assumed heterogeneity in the studies.

BDB and SCM (at n=33) did not show a drastic difference in power and type 1 error. When deciding which method to use in a clinical trial setting, the method that best fits the needs of the specific study should be selected. If a new study was being conducted on pediatric atopic dermatitis, the choice of method should rely on 1) how ethical it is to allocate to the control group, and 2) how easy recruitment is. In cases where it is not ethical to allocate to the control group, a SCM should be used. In cases where recruitment is a challenge, but allocation is not, BDB should be used.